\documentclass[final,5p,times,twocolumn]{elsarticle}

  \usepackage{graphicx}

\usepackage[cmex10]{amsmath}

 \usepackage{amsthm}

\usepackage{xspace}
\usepackage{algorithm}
\usepackage{algpseudocode}
\usepackage{mdwmath}
\usepackage{mdwtab}
\usepackage[caption=false,font=footnotesize]{subfig}
\usepackage{url}
\usepackage{color}
\usepackage{tikz}
\usetikzlibrary{matrix,arrows}
\tikzset{level distance = 4ex, sibling distance = 4ex, every node/.style={fill=gray!30,circle,inner sep=1pt}, baseline=(current bounding box.west)}

\newcommand{\DEF}{
  \mathbin{\smash[t]{\overset{\scriptscriptstyle\mathrm{def}}{=}}}}
\newcommand{\ex}{\ensuremath{\mathbin{\diamond}}\xspace}
\newcommand{\nil}{\ensuremath{\mathord{nil}}\xspace}

\journal{Computers and Security}

\begin{document}

\begin{frontmatter}



\title{Tree-formed Verification Data for Trusted Platforms}


\author[NOV]{Andreas U. Schmidt\corref{cor}}
\ead[url]{http://andreas.schmidt.novalyst.de}
\ead{andreas.schmidt@novalyst.de}
\author[NOV]{Andreas Leicher}
\author[NOV]{Andreas Brett}
\author[IDC]{Yogendra Shah}
\author[ICH]{Inhyok Cha}
\address[NOV]{Novalyst IT AG, Robert-Bosch-Straße 38, 61184 Karben, Germany}
\address[IDC]{InterDigital Communications, LLC,  781 Third Avenue, King of Prussia, PA 19406}
\address[ICH]{Apt 102-202, Joong-Ang Heights, 14-1 Samsung-Dong, Gangnam-Ku, Seoul, Korea}

\cortext[cor]{Corresponding author}

\begin{abstract}
The establishment of trust relationships to a computing platform relies on validation processes.
Validation allows an external entity to build trust in the expected behaviour of the platform based 
on provided evidence of the platform's configuration.
In a process like remote attestation, the 'trusted' platform submits verification 
data created during a start up process.
These data consist of hardware-protected values of platform
configuration registers, containing nested measurement values, e.g., hash values, of loaded or started components.
Commonly, the register values are created in linear order by a hardware-secured operation.
Fine-grained diagnosis of components, based on the linear order of verification
data and associated measurement logs, is not optimal.
We propose a method to use tree-formed verification data to validate a platform. Component measurement values represent leaves, and protected registers represent roots of a hash tree. 
We describe the basic mechanism of validating a platform using tree-formed measurement
logs and root registers and show an logarithmic speed-up for the search of faults.
Secure creation of a tree is possible using a limited
number of hardware-protected registers and a single protected operation.
In this way, the security of tree-formed verification data is maintained.
\end{abstract}


\end{frontmatter}


\section{Introduction}
In a nutshell, the process of building trust in computing platforms follows
a unique pattern~\cite{AUS09B}, common to most architectures of trusted platforms. During start up of the platform, all components
are measured by a protected entity on the platform before they are loaded and executed.
The generation of a chain of trust is an important concept for a Trusted Computing System.
This chain must extend without gaps from system boot up to the current system state,
including all executed instructions and programs.
Every component is required to measure and report the following component before executing it. 
Measurement of the direct successor prevents unmonitored execution of code between measurement 
and actual execution. The start of the measurement process, that is, the taking and recording of measurements, 
is protected by a Root of Trust for Measurement (RoTM), 
which, as a part of the trusted computing base of the platform, is unconditionally trusted, and
may be endowed with security through hardware protection, and trust through certification, for instance.

This is the model process of start up of a trusted platform via a \textit{transitive chain of trust}
according to the Trusted Computing Group (TCG). It comes in two variants:
In~\cite{PCCLIENTBIOS}, \textit{authenticated boot} is specified for PC clients, 
whereas~\cite{MTMREFARC} specifies \textit{secure boot} for mobile platforms. 
The difference is essentially that secure boot adds a local verification and enforcement 
engine that lets components start only if their measurements are equal to trusted reference values.

The mentioned ``measurement'' of system components can be implemented for instance by computing a 
digest value over code and configuration data, e.g., a SHA-1 hash value. 
The measurement values are recorded in a Stored Measurement Log (SML) without particular protection.
However, additional data is compiled from the measurement values by a protected
operation and stored in protected storage, so-called Platform Configuration Registers (PCRs). 
These data identify, after
completion of secure start up, the platform's state uniquely.
PCRs provide information about a system's state with unconditional security.
In particular, they are secure independently of the SML, which, according to TCG standards,
has no particular protection on the platform. PCR values are cryptographically bound to the
contents of the SML by hash-chaining inside a secure environment.
This way to use PCR values to secure the SML
is based on the technique introduced by Schneier and Kelsey for securing audit logs on untrusted 
machines~\cite{1267553,317089}. In fact, it is a simplification, since only the last element 
of the hash chain is kept in a PCR, while the SML normally contains only the measurement values. 
Integrity measurement using the TPM is implemented 
in the Integrity Measurement Architecture (IMA)~\cite{sailer2004design} as a Linux kernel module
to measure the integrity using the TPM and to generate a linear SML.

The main intended use of PCRs and SMLs is validation of the state of a platform by an entity, internal or external.
For this, signed PCR values, may be sent to that entity, the \textit{validator}. The signature over the PCR
values, using Attestation Identity Keys (AIKs) attests to the hardware protection of the PCR value by a TPM.
After following this procedure, defined by the TCG in the attestation protocols~\cite{TCGIWGP1RA}, the 
validator may compare PCR values with a table of known good states, or check the integrity of the SML and 
analyse it entry by entry. It is envisaged by TCG, that validation may eventually be used to take remedial 
steps on trusted platforms, for instance upon first network or service access, as designed in the 
specifications of the Trusted Network Connect working group of the TCG~\cite{TCGTNC}.

The linear, sequential structure of the SML has some known limitations regarding flexibility and efficiency.
In the present paper it is proposed to extend SMLs by data that is organised as a binary tree, more precisely a
Merkle hash-tree~\cite{Merkle1989,10.1109/SP.1980.10006}. We call this a \textit{tree-formed} SML.
Anticipating the following, the main advantages of tree-formed SMLs are as follows:
A logarithmic improvement in the search of failures (components and/or devices which are not in a known good configuration)
allows for efficient verification of many more attached devices. This increases the scalability of
validation to a point that it could be extended to networks rather than just single systems. Also, a tree-formed SML
allows corrupted or compromised components to be found in a hierarchically structured process with a defined granularity.

The paper is organised as follows: To provide a
conceptual background, Section~\ref{sec:line-chain-verif} reviews the linear SML data structure, how it is
generated and used in attestation to validate a platform. An analysis of efficiency and security of the linear SML
serves as reference for later comparison with the proposed tree formed structure. On the basis of that, 
Section~\ref{sec:tree-form-verif} lays out the definition of the tree-formed data structure and its procedural
use for platform validation. Analysis in parallel with that for the linear SML shows the benefits of the
tree-formed approach with regard to efficiency and a more flexible implementation of security through platform
validation. A secure generation algorithm for tree-formed SMLs and associated PCR values, maintaining the security 
assumptions of the linear, standard case is exhibited in Section~\ref{sec:secure-gener-tree}
Section~\ref{sec:implementation} discusses implementation options for tree-formed SML within the scope of existing 
technology. We present a proof-of-concept implementation in an emulated Trusted Computing environment and show 
results of performance experiments. 
The idea of hash trees is a classic, however it has not been used for the validation of platforms, as is shown
in Section~\ref{sec:related-work}, where we position the present contribution in the context of related work.
Section~\ref{sec:disc-outl-concl} concludes the paper with a discussion on applicability of the presented ideas
and outlook to further work.
\section{Linear  Chained Verification Data}\label{sec:line-chain-verif}
Although Trusted Computing is the paradigm we are starting from, we take the opportunity to conceptually abstract
slightly from this concrete case and to introduce a common terminology. This allows a clearer comparison of the
linear chained SML with the proposed tree-formed data structure. Furthermore, we argue that the general concepts
are applicable more broadly than what is in the scope of TCG. 
Examples will be provided in the discussion of Section~\ref{sec:disc-outl-concl}.
\subsection{Data Structure and Generation Procedure}\label{sec:data-struct-gener}
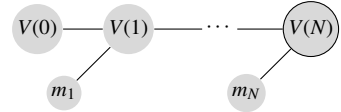
\begin{figure}[t]
	   \centering\footnotesize
	 \begin{tikzpicture}[level distance=12mm,level/.style={sibling distance=32mm/#1}]
	 \node[circle,draw] {$V(N)$}
	child[grow=south west] {node {$m_N$}}
	child[grow=left] {node [style={fill=white}] {$\cdots$}
	  child[grow=left] {node {$V(1)$} 
	    child[grow=left] {node {$V(0)$}}
	    child[grow=south west] {node {$m_1$}}
	  }
	};
	 \end{tikzpicture}
   \caption{Linear Chained SML and PCR generation.}
   \label{fig:linstruct}
 \end{figure}

We call data collected on a platform which is later to be submitted to a validator, and which identifies 
the platform state uniquely collectively \textit{verification data}. The set of measurements, e.g.\ hash values, 
of system components stored in the SML is the standard example of verification data. To elevate the assurance
level of validation, additional, \textit{protected verification data} may be created.
Protected verification data cryptographically protects the bulk of verification
data and is itself protected by some unconditionally trusted functions and resources of the platform, i.e., by
a root of trust. According to TCG, protected verification data is stored in Platform Configuration Registers 
(PCRs, a minimum of $16$ according to version 1.1 of the specification, at least $24$ in version 1.2)
in the TPM, where they can only be accessed by authorised commands. 
TCG specifies computation of protected verification data via the \textit{extend} operation of the 
Trusted Platform Module (TPM, \cite{TPMMAIN}), respectively, the Mobile Trusted Module (MTM, \cite{MTMSPEC})
from the component measurements.
The extend operation builds a linearly ordered, nested chain
of hash values, akin to the Merkle-Damg\r{a}rd transform, as follows:
\begin{equation}
	 V_i(k) \gets  V_i(k-1)\diamond m_k \DEF H \left( V_i(k-1) \|  m_k  \right),
  \label{eq:extend}
\end{equation}
where $V_i$ denotes a \textit{verification data register}, i.e., a particular instance of a piece of protected 
verification data ($i=0,\ldots,23$ for PCRs), $H$ is a collision-resistant
hash function (SHA-1 in case of the TPM), and $m_k=H(\text{data})$ is the $k$-th measurement value taken 
in the start up sequence. This operation is carried out inside the secure execution environment of the 
TPM which is assumed to afford essentially the same protection level as the PCRs themselves.
Figure~\ref{fig:linstruct} shows the linear chained state structure of a verification data register $V$,
created by the extend operation. The intermediate states are transient, i.e., they are normally not recorded
in the SML. Only the final state $V(N)$ is recorded and protected in the register, which is indicated by the 
circle boundary around it.
\subsection{Validation With Linear Verification Data}\label{sec:valid-with-line}
To validate a platform, its protected verification data is sent in an attestation message
to the validator. The attestation message bears a digital signature with a certificate that attests to the 
assurance level provided by the protected verification data. In the case of the TPM, the signature using an
AIK certificate attests to the fact that the signed PCR values have been created and protected by a genuine TPM.
See~\cite{Pirker:2009:PAT:1530327.1530337} for further details.
Alongside with the attestation message, the SML is commonly also submitted as subsidiary data. We call this
data collection \textit{validation data}. The validator is then able,
in principle, to assess the trustworthiness of the platform to any desired granularity,
limited only by the total information conveyed during validation. 

Validation proceeds in an hierarchical manner from attestation message signature verification, over assessment
of protected verification data, to assessment of unprotected verification data:
After verifying the attestation message, the
validator compares protected verification data with \textit{reference data}, i.e., known values of  protected
verification data, e.g., PCR values of known good states of the platform. 
If the protected verification data is found in the validator's library, the platform state is identified
and the validation process ends. If not, the validator needs next to check the integrity of the unprotected 
verification data by retracing the process of creation of the protected verification data, i.e., recalculate
the state $V(N)$ from the measurement values contained in the SML. If that succeeds, the validator may proceed to
look at single entries of the SML to detect \textit{faults} in the system configuration.
\subsubsection{Fault Detection}\label{sec:fault-detection}
The validator may now compare each single
data element of verification data with reference values of single system components, called \textit{Reference Integrity
Measurements}~\cite{MTMREFARC} in TCG parlance. 
If known good reference values are found for each entry in the SML, the validator concludes that the system
state differs from a known good state only by the order sequence in which measurements where entered into the SML.
Assessment of this result may depend on policy.
The opposite case, i.e., one or more entries in the SML differ from their reference values, are what we denote
as \textit{faults} of the corresponding components. 

Detection of faults with linear chained verification data
always requires retracing of a protected verification data value, i.e., it has at least the
computational complexity of $N$ extend operations (SHA-1 operations for PCR values), when the SML contains $N$
measurements. One way to improve efficiency is to extend the library of known reference values of protected 
verification data, but this strategy has limits for complex systems, imposed by combinatorics.
Furthermore, fault detection requires $N$ searches for reference values, equal to a particular one of the SML entries.
\subsubsection{Tamper Detection}\label{sec:tamper-detection}
\textit{Tamper} is anything which breaks the integrity of unprotected verification data with respect to the protected
verification data, i.e., in the linear case, breaks the hash chain. This may or may not be the result of manipulation
by an attacker. The validator may detect tamper only by recalculating the complete chain of states $V(i)$ from the 
submitted $m_k$, $k=1,\ldots,N$ in the SML using (\ref{eq:extend}) and compare the result $V(N)$ with the received protected
verification data register value $V'(N)$. However, the validator cannot even detect the position(s) of tampering in the
SML, i.e., determine the ordered list of original measurement values $[m_k]$ 
which have been replaced by tamper values $[m'_k]$.
Some improvement can be achieved, if also the intermediate states $V(k)$ are transmitted as validation data in an extended
SML. In this case, the validator can retrace the linear chain backwards to find the last position of tamper, i.e., the
largest $k$ for which a received measurement is in $[m'_k]$. 
\section{Tree-Formed Verification Data}\label{sec:tree-form-verif}
The discussion above shows the clear-cut limitations of linear chained verification data: a) Faults can be discovered 
by linear search only, and b) tamper detection is limited to a binary decision.
At best, i.e., in the case that the SML is augmented by all intermediate verification data register states,
determination of the `last point of tamper', i.e., the latest element in the SML which can be proven to
be manipulated, becomes possible. 

Thus, the state-of-the art of verification data, created by linearly chaining extend operations,
is only of limited value for remote diagnostics of a platform, and advanced management
such as component-wise remediation. The space complexity of real world 
SMLs with many hundreds, or thousands, of measured components, makes sifting it through for components
which fail validation, i.e., for which measurement value differs from a ``good'' reference value,
costly.

The point b) also yields
a hint to further improvements by further extensions of validation data. Here we propose one such extension which is
a particular combination of increased capabilities with efficiency, namely to change the topology of the SML  for validation to one or more binary trees with roots protected by verification data registers.
\subsection{Data Structure}\label{sec:data-structure}
Before beginning to construct binary \textit{tree-formed} verification data, we capture the guiding principle, abstracted
from the linear chain example of Section~\ref{sec:line-chain-verif}.
\newtheorem*{val}{Validation Assumption}
\begin{val}
  Platform validation data and process satisfies:
  \begin{enumerate}
  \item[V1] Validation data contains sufficient information to allow for independent, cryptographic proof of the 
assurance level of the integrity of protected verification data.
  \item[V2] Protected verification data contains sufficient information to allow for independent, cryptographic proof
of the same assurance level of the integrity of all verification data.
  \item[V3] Reference data for at least one reference state of a platform for all or parts of verification data 
is available to the validator.
  \end{enumerate}
\end{val}
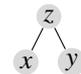
\begin{figure}[t]
\centering
\begin{tikzpicture}
 \node {$z$}
 child {node {$x$}}
 child {node {$y$}};
 \end{tikzpicture}
 \caption{Generic inner nodes in tree-formed verification data.}\label{fig:genericnodes}
 \end{figure}
V1 is just an abstraction of the semantics of the attestation message. V2 and V3 have a more specific meaning
in the present context.
For tree-formed verification data, V2 directly entails the requirement of cryptographic binding of nodes in an
upward direction. A natural implementation of this is by a hash tree in which the generic inner node 
configuration of Figure~\ref{fig:genericnodes}
is related by the operation
\begin{equation}
   z = x\diamond y \DEF H \left( x \|  y  \right).
\end{equation}
To fully satisfy V2, the procedure by which the cryptographic binding is achieved on the platform needs to
satisfy defined security requirements to achieve the desired assurance level. This is symbolised by the use
of `$\diamond$' for the \textit{extend} operation. A concrete procedure to achieve the same assurance level
as for PCRs is described in Section~\ref{sec:secure-gener-tree}. Note that the extend operation is non-commutative, i.e.,
the resulting binary tree structure is left-right ordered, or bi-coloured in mathematical terms.

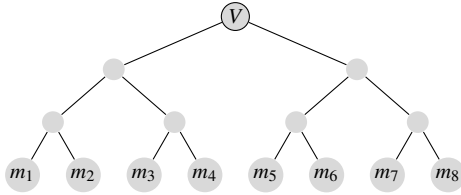
\begin{figure}[t]
 \centering\footnotesize
 \begin{tikzpicture}
[level distance=7mm,
level 1/.style={sibling distance=32mm},
level 2/.style={sibling distance=16mm},
level 3/.style={sibling distance=8mm}]
 \node[circle,draw] {$V$}
 child { 
   node {\phantom{x}}
   child {
     node {\phantom{x}} 
     child {node {$m_1$}} 
     child {node {$m_2$}}
   }
   child {
     node {\phantom{x}}
     child {node {$m_3$}}
     child {node {$m_4$}}
   }
 }
 child { 
   node {\phantom{x}}
   child  {
     node {\phantom{x}} 
     child {node {$m_5$}} 
     child {node {$m_6$}}
   }
   child {
     node {\phantom{x}} 
     child {node {$m_7$}} 
     child {node {$m_8$}}
   }
 }; 
 \end{tikzpicture}  
   \caption{Generic tree-formed verification data of depth 3.}
   \label{fig:generic-tree}
 \end{figure}

Figure~\ref{fig:generic-tree} shows the generic structure of a tree-formed SML of depth $d=3$ 
with component measurements at
the leaves and a protected verification data register $V$ at the root. 
In this example, the number of measurements is a power of $2$. Section~\ref{sec:secure-gener-tree} describes 
a method to ``fill up'' a tree-formed SML when this is not the case.

The basic concept of tree-formed verification data readily extends to $n$-ary trees and even general
trees of non-uniform arity. However, at every node with arity higher than $2$, some loss of granularity 
of semantics will occur, since the $n$ children at such a node form a linear hash chain of verification data,
incurring the drawbacks described in Section~\ref{sec:valid-with-line}.
\subsection{Validation With Tree-Formed Verification Data}\label{sec:validation-with-tree}
The validation strategy for tree-formed verification data described in this section exploits all available
information at every tree node. 
The discussion proceeds in parallel with the linear chained case for easy comparison.
In Section~\ref{sec:efficiency}, the average computational cost is calculated
in relation to the number, respectively, relative share of failed measurements.
First, fault detection is described in the process of \textit{diagnostic validation} in
Section~\ref{sec:fault-detection-1}. Section~\ref{sec:tamper-detection-1} describes the extended 
capabilities of tree-formed data for tamper detection. Section~\ref{sec:efficiency} presents
quantitative performance estimates for diagnostic validation for a simple statistical model 
of fault distribution.
\subsubsection{Fault Detection}\label{sec:fault-detection-1}
\begin{figure}[b]
 \begin{gather*}
 \begin{tikzpicture}
 \node {$g$}
 child {node {$g$}}
 child {node {$g$}};
 \end{tikzpicture}\tag{a}\\
 \begin{tikzpicture}
 \node {$b$}
 child {node {$b$}}
 child {node {$g$}};
 \end{tikzpicture}\ 
 \begin{tikzpicture}
 \node {$b$}
 child {node {$g$}}
 child {node {$b$}};
 \end{tikzpicture}\ 
 \begin{tikzpicture}
 \node {$b$}
 child {node {$b$}}
 child {node {$b$}};
 \end{tikzpicture}\tag{b}\\
  \end{gather*}
 \caption{Classification of node configurations relating to fault 
detection with tree-formed verification data.}\label{fig:nodeclassfault}
 \end{figure}
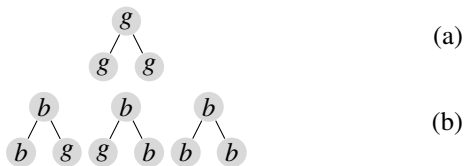
The aim of diagnostic validation  is to find the subset of fault leaves.
For the concrete instantiation of V3 for diagnostic validation,
we assume that the validator uses a complete reference tree for comparison of inner nodes and leaves.
Then, validation can start from the root of the tree, i.e., a protected verification data 
register value $V$, traversing the tree downward.
This yields the leaf set of components for which measurements differ from reference values.
In traversing the tree, a depth-first search with pruning is applied, and decisions are taken
at every branching node.
The tree values at a node and its two children are
compared with the reference tree values of the same node
positions, and the results are noted as $g$ (good) for agreement and $b$ (bad) for discrepancy.
In this notation, the situations shown in Figure~\ref{fig:nodeclassfault} can occur (in the absence of 
tamper as assumed in this subsection).

In case (a), the whole subtree below this parent node is validated positively,
and traversal ends at this node.
In the cases (b), the parent node is recalculated by the validator applying
the extend operation to the child node values, to check the integrity of the parent node.
Validation proceeds to the next tree level, traversing the
subtrees where bad values are found, i.e., left, right, or both subtrees
in (b), respectively. The unique property of diagnostic validation 
is that it \textit{finds all faults with untampered measurement values in untampered subtrees}. 
\subsubsection{Tamper Detection}\label{sec:tamper-detection-1}
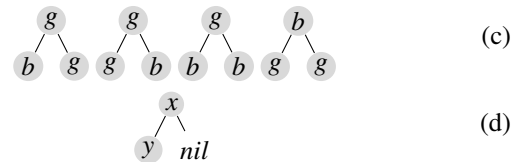
\begin{figure}[b]
 \begin{gather*}
 \begin{tikzpicture}
 \node {$g$}
 child {node {$b$}}
 child {node {$g$}};
 \end{tikzpicture}\ 
 \begin{tikzpicture}
 \node {$g$}
 child {node {$g$}}
 child {node {$b$}};
 \end{tikzpicture}\ 
 \begin{tikzpicture}
 \node {$g$}
 child {node {$b$}}
 child {node {$b$}};
 \end{tikzpicture}\ 
 \begin{tikzpicture}
 \node {$b$}
 child {node {$g$}}
 child {node {$g$}};
 \end{tikzpicture}\tag{c}\\
 \begin{tikzpicture}
 \node {$x$}
 child {node {$y$}}
 child {node [style={fill=white}] {\textit{nil}}};
 \end{tikzpicture}\tag{d}
 \end{gather*}
 \caption{Classification of node configurations relating to tamper 
detection with tree-formed verification data.}\label{fig:nodeclasstamper}
 \end{figure}
If, during diagnostic validation, the  value of a node recalculated from its 
children does not match the value at the parent node,
this indicates a tamper in one of the subtrees of which those children are the
roots. Which node, or set of nodes, was actually tampered with, cannot be determined with certainty. Therefore, tamper detection is analogous to the best-effort restricted localisation of 
(the last) tamper position for the linear chain (extended by intermediate verification
data register states), which was described in Section~\ref{sec:tamper-detection}.

However, for tree-formed verification data, there are more cases of tamper detection,
which occur in the configurations Fig.~\ref{fig:nodeclasstamper}~(c) 
of nodes shown in Figure~\ref{fig:nodeclasstamper}.
It should be noted that this detection takes place without recalculating 
an extend operation, because the shown triplets have only one bad value, which cannot occur.
The last situation, Fig.~\ref{fig:nodeclasstamper}~(d), only occurs when the binary tree is incomplete,
and a right branch is null. Then, as a convention, value $x$ must equal value $y$, in which
case traversal proceeds to the left, and otherwise tamper is detected at this node.
 
Let us compare the impact of tamper on fault detection for the linear and tree-formed case.
Since tamper in the linear SML cannot be localised, a failure to
reproduce the PCR value also  means that diagnostic validation becomes impossible, and failed
components cannot be distinguished from good ones.
For tree-formed verification data, the situation is much better. If a subtree is identified, where manipulation
of the SML is suspected, the complement of it in the verification data tree can still be validated for
fault detection. 
\subsubsection{Efficiency}\label{sec:efficiency}
For diagnostic validation, a significant speed-up in determining the set
of failed components, may be expected by the binary search strategy.
A principal advantage of validating tree-formed SMLs is that subtrees with
a correct root can be discarded from further search for failed components. 
In this section we lay out a simple, probabilistic
model to quantitatively assess the performance of tree validation.
We assume that recalculating the extend (hash) operations is the dominant cost
factor to estimate validation complexity, while comparisons are cheap.
Assume a random, i.i.d.\ set of fault leave measurement values. 

When visiting an inner parent node which differs from the corresponding node in the reference tree, i.e.,
a bad parent node, one of the situations in Fig.~\ref{fig:nodeclassfault}~(b), or the rightmost 
configuration of Fig.~\ref{fig:nodeclasstamper}~(c) is encountered. 
In the latter case, no recalculation of the parent node
needs to be performed since tamper is directly detected. 
The subtree with configuration Fig.~\ref{fig:nodeclasstamper}~(c)
is discarded from further traversal, since it cannot be assumed to yield trustworthy information
about faults. In this case, further steps depend on the validator's policy.
The node configurations Fig.~\ref{fig:nodeclassfault}~(b) are the ones which require re-calculation of the
parents from the children by $\ex$, to confirm that the configuration,
which is unknown from the validator's reference tree, is authentic.
The subtrees whose roots are good children of the bad parent node under scrutiny,
are discarded from further traversal.
Note that this procedure implicitly
excludes the subtrees with root configuration Fig.~\ref{fig:nodeclassfault}~(a) and the three left configurations of 
Fig.~\ref{fig:nodeclasstamper}~(c) from further traversal. 
They correspond to ``known good'' subtrees with untampered roots.
These subtrees may be considered in further forensic evaluation
of the SML tree, wherever this makes sense.

 \begin{figure}[t]
\centerline{\resizebox{0.5\textwidth}{!}{\includegraphics{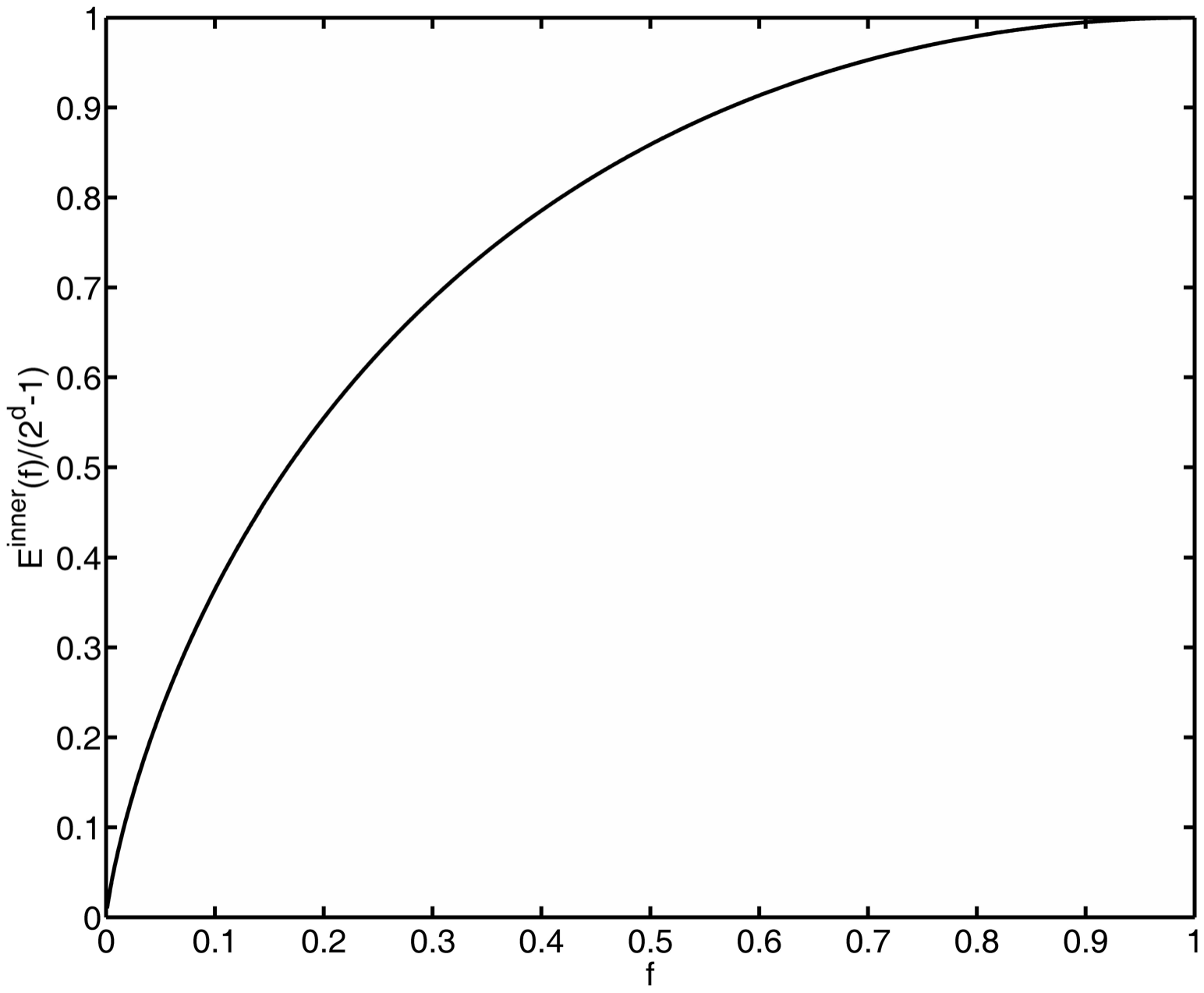}}} 
 \caption{Expected fraction of bad inner nodes on random distribution of $2^df$ bad leaves for $d=16$.}
 \label{fig:bad_fraction}
 \end{figure}
Summarising, we see that diagnostic validation requires to visit and perform an extend operation
at all bad inner nodes in the union of all paths from failed (bad) leaves to the root, i.e.,
one hash operation for each bad node which is not directly recognised as tampered by being in
the rightmost configuration of Fig.~\ref{fig:nodeclasstamper}~(c).
Therefore it is sufficient for a first-best 
complexity analysis to assume a tamper-free, full tree
of depth $d$, and 
a subset of i.i.d.\ bad leaves constituting a fraction $f\in[0,1]$ of all leaves.
The number of bad leaves is $2^d f$. 
The cost of diagnostic validation is dominated by the expected number $E^{\text{inner}}(f)$ of 
bad inner nodes. This number can be calculated
as shown in Section~\ref{sec:calc-expect-numb}.

Figure~\ref{fig:bad_fraction} shows the fraction  
of the $2^d-1$ total inner nodes, for $d=16$, at which a hash operation
will occur under the assumptions above.
This represents the number of hash operations which are necessary to determine the bad components with certainty.
The reference case of a linear SML requires $2^d$ hash operations to recalculate the final PCR value.
This case is roughly represented by the upper ordinate axis of Figure~\ref{fig:bad_fraction}.

When the computational cost of comparisons to reference values is taken into account, 
the situation is slightly different.
Tree traversal for diagnostic validation descends along the bad inner nodes which fail comparison
with the reference tree's corresponding inner node. For that, both children of a bad inner
node have to be compared in every case, so that the complexity in terms of comparisons
is twice the number $E^{\text{inner}}(f)$. The linear SML requires all $2^d$ measurements 
to be compared with reference values.

If $h$ is the cost of a hash operation at the validator, and $c$ the cost of a comparison
of two hash values (160 Bit for SHA-1), then the total validation cost of the linear case is
$2^d(h+c)$. This is the least effort to obtain the same information
from a linear SML as by diagnostic validation of a tree-formed SML.
For the tree-formed SML on the other hand (including the root in the count), 
the cost is $(E^{\text{inner}}(f)+1)(2c+h)$. Tree-formed diagnostic validation is more efficient than
validation of the linear chain if
\[
\frac{E^{\text{inner}}(f)+1}{2^d}\leq \frac{h+c}{h+2c} = \frac{\lambda+1}{2\lambda+1},
\]
where $\lambda=c/h\ll 1$. With a very generous margin, $\lambda<0.01$, which yields a bound of $0.99$ for
the r.h.s. Then, for $d=16$, tree-formed validation is expected to be more efficient for
fractions $f$ of bad leaves as high as $85\%$.

We see that diagnostic validation of tree-formed verification data always
performs better in terms of hash operations than with a linear chain, 
and outmatches the linear chain even for large fractions
of bad components, under reasonable assumptions,
and becomes vastly advantageous for small fractions of failed
components. 

It can be expected that tree validation is yet more efficient when the bad leaves are non-uniformly
distributed, e.g., exhibit clustering. 
While we have directly compared linear and diagnostic tree validation, it should be noted that
linear validation becomes impossible if the recalculation of the final PCR fails, since then,
comparison of single measurements does not yield reliable information --- each measurement could
be faked in the SML to hide the one which broke the hash chain. In conclusion, the principal, semantic
advantage of tree-formed validation data comes about even at decreased computational complexity
for the validator.
\subsection{Extensions for Efficiency and Augmented Functionality}\label{sec:extens-effic-augm}
Unlike the linear chain, tree-formed verification data has a natural hierarchical structure, namely \textit{subtrees}
of a verification data tree, which opens paths for increased functionality. A particular problem of validation
(linear and tree-formed) is combinatorial: It may be difficult in practice to keep reference data
of a whole platform. Such data may be large and depend sensitively on platform configuration, for instance load orders.
In fact, since the Merkle-Damgard transform used for the extend operation is neither commutative nor associative 
(unlike some proposed \textit{multi-set hash functions}~\cite{Clarke03incrementalmultiset}),
the PCR values of linear chains as well as the tree root values depend on the exact measurement sequence.
This is a principal, even desired from a security viewpoint, feature of this transformation, 
reflecting the ability to detect the smallest changes in input data according to assumption V2. 
This information-preserving feature of the Merkle-Damgard transform entails, however, 
complexity in the interpretation of the contained information, in particular, 
combinatorial growth of reference data bases. 

As one approach to the problem, the validator could keep a database of known good \textit{subtree} root values of 
subtrees of various depths, instead of a large number of \textit{complete} reference trees. 
These values may be stored as pairs $(r,h)$, where $r$ is the node value of the subtree 
root and $h$ is the depth of the subtree. The depth $h$ ranges from the depth $d$ of the full SML tree to $0$, 
where $d=0$ means that the corresponding reference value is actually a leaf. When reaching a certain level 
$\ell$ in diagnostic tree traversal, the validator compares, by breadth-first search, 
the node values of the received SML tree at this level to the known good values in the level set
\begin{equation}
  \label{eq:level-set}
  \{(r,h)\mid h=d-\ell\}. 
\end{equation}
Every subtree in the received tree-formed verification data with a root equal to one of the elements in the
level set can then be discarded from further traversal. The validator could as well apply different search
strategies using the subtree root database, e.g., look for known good subtree roots in the received verification
data tree first, and then verify them along their paths to the full tree's root.

Practically, subtree roots would be chosen such that they correspond to subsystems of a system, for which the start
up sequence is less likely to change than for the whole platform. Further extensions are rather abundant.
For instance, the validator may dynamically feed the reference tree database by interactive learning based 
on previous platform validation results. One possibility is to perform diagnostic validation the first time a 
platform validates, and generate known good subtrees from subsequences of leaf measurements which correspond 
to a good platform configuration according to the validator's component (i.e. leaf) reference values and policy. 
When more and more platforms validate, the validator can then build a statistically weighted database of known 
good subtrees and platform configurations according to criteria such as commonality and risk (associated to 
configurations). Frequently appearing subtrees may be searched for first, for instance.

The full power of using subtrees can be unleashed by combining them with well-known concepts of certification
by trusted third parties (TTP). Platforms may obtain digital certificates for subtree roots from such TTPs, which
may for instance represent certification authorities for software modules. Such subtree certificates may be included
in a verification data tree supplanting the corresponding subtrees. These concepts are further described in~\cite{DBLP:TFV2}.

Finally, we cursorily touch on some generic (i.e., applying to trees and linear chains alike) methods to 
alleviate the combinatorial problem.
\begin{itemize}
\item \textit{Pre-} or \textit{post-ordering} of the measurement sequence during start-up. Either, a known order is applied before the
measurements are effected (pre-ordering), which results in a rigged formation of verification data, 
or the measurement process is left untouched and ordering is applied later, on the completed verification data, in some 
secure way (post-ordering). Both approaches may incur security and practical problems.
\item Communicating sequencing to the validator. The platform may transmit a sequence of component identifiers, 
indicating the sequencing of all or part of the measurements. In the case of trees this may already help
the validator to find reference values of subtrees, since certain subsequences of identifiers may be used as keys to
the subtree root database.
\item Application of sort criteria, such as lexicographic, fault probability, etc., to the methods of pre- or post-ordering.
\end{itemize}
\section{Secure Generation of Tree-Formed Verification Data}\label{sec:secure-gener-tree}
In this section, we show a practical solution for the problem of creating tree-formed verification data
maintaining assumption V2, using only a limited number of protected verification
data registers to securely generate one root verification value.

Secure creation of verification data which represents root nodes of hash trees poses 
a particular problem. In the normal extend operation of a TPM used for linear chaining, 
only the measurement value taken by
the RoTM on a component, and the current PCR value are used, 
and the operation itself is carried out in the hardware
protected TPM. Thus, in particular, previous measurements
already processed must not be used in the
SML generation process,
since they would have to be fetched from the SML which is not assumed to be protected.
This would compromise assumption V2.

For tree-formed verification data representing a hash tree, where 
adding a new leaf always affects $d-2$ inner nodes of the tree, where $d$ is the tree's depth,
the challenge is to generate tree-formed verification data, from sequentially arriving
measurement values, exclusively inside a limited number
of hardware protected registers (PCRs), using only a single leaf measurement value as input.
Concurrently we are looking for a compact algorithm which lends itself to implementation
on secure hardware and would potentially require only minimal changes on existing TPM
architecture.

It should be noted that every reference to the concrete embodiments of Trusted Computing specified
by the TCG made in this paper, in particular TPM operations, PCRs, and SML, are examples for possible
realisations of the presented concepts. The algorithms and procedures can in principle be applied
to every security technology with the minimum capabilities which are used by them (discussed
in the Section~\ref{sec:compl-tree-form} on the complexity of the procedure).
\subsection{Tree Formation Procedure}\label{sec:tree-form-proc}
%
%
In our proposed solution, one of the hardware protected registers 
$\mathcal{V}\DEF\{V_1,\ldots,V_r\}$, 
e.g., PCRs, contains the root of the final tree.
The tree is chosen to be binary, to keep the algorithm as compact as possible and to provide 
a fine grained detection of failed components.
The leaves are carrying the measurement values, while the inner nodes are stored in a modified SML.
The SML is modified in a way to support the tree structure of the validation data, i.e. it is no longer a
linear list of measurement values but the data structure must support standard tree operations and traversals.
For efficient search during platform validation, the SML must support the addition of 
new leaves and retain edge relations.
Adding a new measurement at a leaf to the tree at depth $d$ requires recalculation 
of all $d-1$ inner nodes of the leaf's reduced hash tree and the tree root which is stored in a $V\in\mathcal{V}$.
A Merkle tree has a natural colouring of edges as ``left'', respectively, ``right'' ones,
since the binary extend operation~\eqref{eq:extend}, is non-commutative.
Leaves inherit this order and are added from left to right.
The binary, $d$-digit representation of leaf $n$, $0\leq n \leq 2^d-1$, 
denoted by $\langle n \rangle$, yields natural coordinates for the
inner nodes and edges on the unique path from  leaf to  root.
That is, the $k$-th digit (counted from the MSB, $k=1,\ldots, d$), 
$\langle n \rangle_k$, determines whether the node at depth $k-1$ on this path is 
connected by a left, respectively, a right edge, 
by $\langle n \rangle_k=0$, or, $\langle n \rangle_k=1$,  respectively.

We make the following assumptions:
(1) the root of every subtree created during the execution of the 
algorithm must always be stored securely in a $V\in\mathcal{V}$.
(2) If two subtrees (measurement values are subtrees of depth 0) with the 
same depth $d'$ exist, they can be merged to a single tree of depth $d'+1$.
(3) The merge operation must preserve assumption (1), i.e., 
one of the two $V$ protecting the roots of the subtrees is freed after the merge operation.
Using these assumptions, the update algorithm for a newly arriving measurement
value can be formulated such that registers $V_1,\ldots,V_{d-1}$ always contain
the current state of ``active'' subtrees of depth $d,\ldots, 1$, and in particular $V_d$
is affected by the incoming measurement values of components.
``Active'' here means a subtree the root of which awaits completion by merging
its two completed child subtrees.
Care is taken in the formulation so that only the actual measurement value,
protected registers, and the normal extend operation are used,
and no unprotected memory places are involved. 
Denote an empty node in the full binary tree of depth $d$ by \textit{nil}.
The tree formation is performed by Algorithm~\ref{alg:tree-form}.
 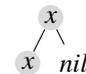
\begin{figure}[b]
   \centering
 \begin{tikzpicture}
 \node {$x$}
 child {node {$x$}}
 child {node [style={fill=white}] {\textit{nil}}};
 \end{tikzpicture}
   \caption{Default configuration of nodes at the right edge of a tree-formed SML.}
   \label{fig:defaultconfig}
 \end{figure}

If $n<2^d$, the tree is incomplete at the right edge,
and the cleanup procedure shown in Algorithm~\ref{alg:tree-clean} is then needed.
Algorithm~\ref{alg:tree-clean} results in
a final merge of roots such that $V_1$ ultimately contains 
all subtree information. Note that this cleanup procedure is
only reached if the tree is not already full, due to the test
in lines~\ref{ln:fulltestbegin}--\ref{ln:fulltestend} of 
algorithm~\ref{alg:tree-form}. The rule by which the tree
is completed is that the configuration shown in Figure~\ref{fig:defaultconfig}
is correct at the right edge.
All non-\nil inner nodes are written to the SML, even if they are
the result of forwarding along a left edge (entailing
minor redundancy).
Formally, the above rule corresponds may be interpreted as modifying the notion of the `\ex' operation
such that $x\ex\nil=x$, as explained in \ref{sec:useful-convention}.

\begin{algorithm}[!t]
\caption{Tree formation algorithm}\label{alg:tree-form}
\begin{algorithmic}[1]
\Require $V_1,\ldots,V_d\in\mathcal{V}$, $m \in \{0,1\}^{× 160}$
\Ensure $V_1,\ldots,V_d=\textit{nil}$ \Comment Initialise subtree roots empty.
\State $n \gets 0$
\While{$(m \gets \text{RoTM}) \neq \textit{nil}$}  \Comment Get new measurement.
\State $m \to \text{SML}$ \Comment  If non-empty, add as new leaf.
\If{$\langle n \rangle_d=1$} \Comment A value arriving from right \label{ln:test1}
  \State $V_d \gets V_d \diamond m$ \Comment extends the first inner node, 
  \State $V_d \to \text{SML}$ \Comment which is purged to the SML.
\State $k\gets d-1$ \Comment Update subtrees of depth $2,\ldots$,
\While{$(\langle n \rangle_k=1) \land (k>0)$} \Comment while coming from right\label{ln:test2}
\State $V_k \gets V_k \diamond V_{k+1}$
\State $V_k \to \text{SML}$
\State $k\gets k-1$
\EndWhile
\If{$k=0$} \label{ln:fulltestbegin}
\State \textbf{return} ``tree full''
\EndIf \label{ln:fulltestend}
\State $V_k \gets V_{k+1}$ \label{ln:shifta}
\Else \Comment If it is arriving from the left,
  \State $V_d \gets m$ \Comment updates the first inner node,\label{ln:M}
\EndIf
\State $n \gets n+1$ \label{ln:cont}
\EndWhile
\algstore{brk}
\end{algorithmic}
\end{algorithm}
\begin{algorithm}[!t]
\caption{Cleanup of an incomplete tree}\label{alg:tree-clean}
\begin{algorithmic}[1]
\algrestore{brk}
\For{$k\gets k-1\textbf{ to }1$}
\If{$\langle n \rangle_k=1$}
\State $V_k\gets V_k\diamond V_{k+1}$
\State $V_k\to$ SML
\Else
\State $V_k\gets V_{k+1}$  \label{ln:shiftb}
\State $V_k\to$ SML
\EndIf
\EndFor
\end{algorithmic}
\end{algorithm}

\begin{figure}[b]
   \centering
 \begin{tikzpicture}[level distance=6mm,level/.style={sibling distance=32mm/#1}]
 \node[circle,draw]{$12$}
 child { 
   node {$7$}
   child {
     node {$3$} 
     child {node {$1$}} 
     child {node {$2$}}
   }
   child {
     node {$6$} 
     child {node {$4$}} 
     child {node {$5$}}
   }
 }
 child { 
   node [dashed,draw,fill=green] {$11$}
   child {
     node [dashed,draw,fill=green] {$10$} 
     child {node {$8$}} 
     child {node {$9$}}
   }
   child {
     node [style={fill=white}] {\textit{nil}} 
     child {node [style={fill=white}] {\textit{nil}}} 
     child {node [style={fill=white}] {\textit{nil}}}
   }
 }; 
 \end{tikzpicture}  
   \caption{Order of nodes in a tree-formed SML.}
   \label{fig:node-order}
 \end{figure}
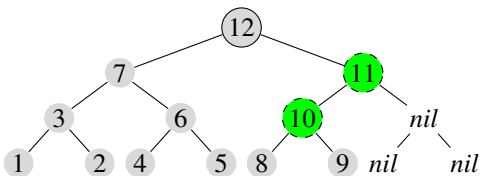

It is interesting to note that, if leaves and 
inner nodes are appended to the SML in the order prescribed
by Algorithm~\ref{alg:tree-form}, a natural
serialisation of the resulting tree is obtained. 
This order is shown in Figure~\ref{fig:node-order} for an incomplete tree of depth 3.
The marked entries $10$ and $11$ in the resulting SML are identical, since $11$ is
created by a forward operation of the cleanup algorithm~\ref{alg:tree-clean}.
The cleanup algorithm writes nodes 10, 11, and finally the root 12 to the SML. Note that the leaf 9
has already been written to the SML in the last run of Algorithm~\ref{alg:tree-form}.

The SML order can be used to address tree nodes in the SML by a binary search. 
Given a sequence number $K$ in the SML of length $2^{d+1}-1$, such a search proceeds from the root,
which is the last entry. The remaining $2^{d+1}-2$ entries are equally partitioned
into portions of size $2^{d}-1$,
and it is decided if $K$ is in the left or right part. This procedure is iterated 
until $K$ points to the rightmost element in the current part.
The sequence of decisions made yields the sequence of left-right edges leading from the
root to the node with index $K$ in the SML.

The tree-formation algorithm can easily be adapted to trees of arbitrary, uniform, arity, say $b$.
For this, the binary coordinate $\langle n \rangle$ has to be replaced by the $b$-ary coordinate $\langle n \rangle^{(b)}$
and its $d$-th, respectively, $k$-th digit evaluated in line~\ref{ln:test1}, respectively,~\ref{ln:test2} of
algorithm~\ref{alg:tree-form}, where the evaluated expression has to be changed to $\langle n \rangle^{(b)}_d=b-1$, respectively,
$\langle n \rangle_k=b-1$. Algorithm~\ref{alg:tree-clean} has to be adapted accordingly.
A further generalisation to arbitrary trees requires only establishment of the associated node coordinates,
i.e., of the mapping $n \to \text{node}$.
\subsection{Maximum Tree Capacity}\label{sec:maxim-tree-capac}
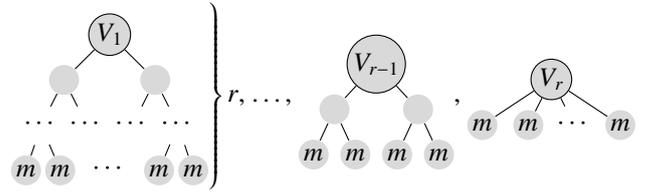
\begin{figure}[ht]
   \centering
 $\left.
 \begin{tikzpicture}[level distance=6mm,level/.style={sibling distance=12mm/#1}]
 \node[circle,draw] {$V_1$}
 child [sibling distance=6mm] {
   node {\phantom{$m$}}
   child {
     node [style={fill=white}] {$\dotsb$}
     child {node {$m$}} 
     child [sibling distance=5mm] {node {$m$}}
   } 
   child {
     node [style={fill=white}] {$\dotsb$}
   }
 }
 child [white, sibling distance=6mm] { child { child { node [style={black, fill=white}] {$\dotsb$} } } }
 child [sibling distance=6mm]{
   node {\phantom{$m$}}
   child {
     node [style={fill=white}] {$\dotsb$}
   }
   child {
     node [style={fill=white}] {$\dotsb$}
     child [sibling distance=5mm] {node {$m$}} 
     child {node {$m$}} 
   }
 };
 \end{tikzpicture}
 \right\}r$, $\dotsc$,
 \begin{tikzpicture}[level distance=6mm,level/.style={sibling distance=11mm/#1}]
 \node [circle,draw] {$V_{r-1}$}
 child {node {\phantom{$m$}} child {node {$m$}} child {node {$m$}}} 
 child {node {\phantom{$m$}} child {node {$m$}} child {node {$m$}}};
 \end{tikzpicture},
 \begin{tikzpicture}
 \node [circle,draw] {$V_r$}
 child {node {$m$}}
 child {node {$m$}}
 child {node [style={fill=white}] {$\dotsb$}}
 child {node {$m$}};
 \end{tikzpicture}
   \caption{Maximum capacity arrangement of tree verification data. Measurement values at the leaves are 
 indicated as $m$.}
   \label{fig:max-cap-trees}
 \end{figure}

%
%
It is clear from the generation procedure that, with a limited number, $V_1,\ldots, V_r$, 
of verification data registers, only a finite number of components at the leaves of trees 
can be covered. In contrast, the hash chain created by the standard, linear extend, ending in a single PCR
value, is in principle of unlimited length.
The maximum capacity of trees generated with $r$ root registers can be calculated as follows.
The procedure for the first register, $V_1$, can use the $r-1$ other registers as
a pipeline of length $r-1$ to build a tree of depth $r$. When $V_1$ is occupied,
the second register can support a tree of depth $r-1$, and so on, until the last
register, $V_r$, for which the pipeline has length $0$ and the tree depth $1$.
Thus the total number of leaves carried by the trees of all registers is
\begin{equation}
  \label{eq:capacity}
  C_\text{trees} = \sum_{k=1}^{r} 2^k = 2^{r+1}-2.
\end{equation}
For $r=24$, the number of PCRs of a TPM adherent to the v~1.2 specification, 
this yields $33,554,430$ places for component measurements at the leaves of
the $r$ trees. If restricted to the last 16 PCRs, since, for instance, according
to the PC Client specification of the TCG~\cite{PCCLIENTBIOS} PCRs 0--7 are reserved,
(\ref{eq:capacity}) still counts $131,070$ measurements (see Section~\ref{sec:implementation} 
for a discussion of implementation issues with standard TPMs).
Since the number of measurements to be taken during start up or at run-time
is not \textit{a priori} known, the last register can, as a fallback, be linearly extended
after the capacity limit is reached. Figure~\ref{fig:max-cap-trees} shows this arrangement.
\subsection{Complexity of Tree Formation}\label{sec:compl-tree-form}
%
%
The spatial complexity of the tree formation algorithm is very small.
As internal data needs precisely three locations: 
$d\in\{1,\ldots,r\}$, $n\in\{0,\ldots,2^d-1\}$, and $k\in\{1,\ldots,d\}$,
the size of that data is at most $d + 2\lceil\log_2 d\rceil\leq r + 2\lceil\log_2 r\rceil$ Bits.
Additionally, depending on implementation one register may be required
to receive and hold the current measurement value, and/or as an intermediate register
for the operations on verification data registers. 
The SML increases moderately in size. For a completely filled binary tree of depth $d$,
$2^{d+1}-2$ node values, including leaf measurements, are stored in the SML (the root node
is contained in a $V_i$). That is, the tree-formed SML is less than double the size of the
linearly formed SML containing only measurement values.

For an estimation of the temporal complexity, we consider a full tree of depth $d$, i.e.,
$2^d$ leaf measurements.
The various operations involved in algorithm~\ref{alg:tree-form} are
\begin{itemize}
\item[$M$] Add measurement to $V_d$; $V_d \gets m$.
\item[$S_V$] Store a verification data register to SML; $V_k\to$ SML.
\item[$S_m$] Store measurement to SML; $m\to$ SML.
\item[$V$] Copy verification data register; $V_k\gets V_{k+1}$.
\item[$E_1$] Extend $V_d$ with measurement; $ V_d\gets V_d\diamond m$.
\item[$E_2$] Extend inner node registers; $V_k\gets V_k\diamond V_{k+1}$.
\end{itemize}
The symbols introduced in the list items above denote the execution times of the described operations.
The one missing operation $m\gets$ RoTM can be subsumed in $S_m$.

By the structure of the tree, the occurrences of the operations are easily counted.
$S_m$ occurs at each leaf, i.e, $2^d$ times.
$E_1$ and $M$ occur at each inner node at depth $d-1$, i.e., $2^{d-1}$ times.
$V$ and $E_2$ occur at each inner node from depth $d-2$ upward, i.e., $2^{d-1}-1$ times.
Finally, $S_V$ occurs at each inner node of the tree except the root, which remains
in $V_1$. That is, $S_V$ occurs $2^d-2$ times. Altogether this yields the estimate
\[
2^{d-1}(E_1+M)+(2^{d-1}-1)(V+E_2)+2^d S_m + (2^d-2)S_V
\]
for the algorithm's execution time, disregarding flow control.
Grouping similar operations $\{E_1, E_2\}$, $\{M, S_V, S_m\}$ yields
\[
2^{d-1}(E_1+E_2)-E_2+2^{d-1}(M+2S_V+2S_m) -2S_V+(2^{d-1}-1)V.
\]
Assuming that all memory operations are approximately equally time-consuming
and bounded by a common constant
\[
M\approx S_V \approx \textstyle\frac{1}{2} S_m \approx \textstyle\frac{1}{2} V \leq S, 
\]
(where a factor $2$ is included in $V$ for a na\"ive read/store implementation, and in
$S_m$ for the missing operation mentioned above), and likewise for the extend operations
\[
E_1\approx E_2 \leq E.
\]
Then, a coarse estimate for the temporal complexity of tree formation for $d>1$ is
\[
\leq 2^d\left( E+4\textstyle\frac12S \right) - (E+4S).
\]
When extend operations are the dominating factor, it is interesting to note
that tree formation actually needs one extend operation \textit{less} than
the linear chain of authenticated boot.
\section{Implementation}\label{sec:implementation}
This section describes implementation options and a concrete proof-of-concept implementation 
for the generation of tree-formed verification data according to Section~\ref{sec:secure-gener-tree}
within the reference framework provided by Trusted Computing.

With regard to the tree-formation algorithm itself, to achieve the same level of security
as TCG standard compliant trusted boot processes, all operations on verification data registers
should run inside the hardware-protected TPM environment. 
Most operations of the tree-formation algorithm listed in Section~\ref{sec:compl-tree-form} are,
however, are not standard TPM functions.
In fact, only the normal extend operation $E_1$ is an internal standard function, and
$S_V$ and $S_m$ can be realised by PCR read operations. 

We first discuss the minimal modifications that would be necessary to extend a TPM to
turn PCRs into tree-formed verification data registers, while the tree-formation
algorithm may still run outside the TPM. 
Then, we propose a new TPM-internal command for tree formation\footnote{%
A third variant, which is not further discussed here is a software-based implementation
of tree-formed verification data, where the root registers are soft registers managed by
a trusted application, and where the current state of such registers is protected by a `real' 
register, e.g., a PCR.}.
Finally, we explain the experimental implementation
of tree-formation with the TPM emulator~\cite{Strasser2004} integrated in the TPM software 
emulation environment ethemba~\cite{ethemba}.
\subsection{Minimal TPM Modifications for Tree-Formation}\label{sec:minim-tpm-modif}
Let us first take a minimalist approach to implementing tree-formation and carve out
the least changes to a standard TPM that would enable PCRs for use with
the algorithms~\ref{alg:tree-form} and~\ref{alg:tree-clean}. This requires implementing only the
elementary operations listed in section~\ref{sec:compl-tree-form} by TPM commands or
modifications thereof. The core of the algorithm, including the bookkeeping tasks
on registers representing inner nodes' current states, could then be realised as
a software root of trust for performing tree-formation in a system integrity
measurement process such as authenticated or secure boot. 
The soft RoTM for tree formation can for example be realised in a late-launch
compartment using for instance Intel's TXT or ARM TrustZone technologies.

The operations $S_V$ and $S_m$ pose no problem and can be realised by \texttt{TPM\_PCRRead} commands
or directly in the tree formation software, respectively.
$E_1$ occurs at every right edge at the lowest level of the tree, and extends a $V$ already containing
a measurement value which came from the left sibling of the measurement which is
extended into $V$. Therefore, $E_1$ is precisely the standard \texttt{TPM\_Extend} operation
defined by (\ref{eq:extend}). $E_2$ also occurs at right edges inside the tree and,
in turn, is straightforwardly modelled by \texttt{TPM\_PCRRead} followed
by a \texttt{TPM\_Extend}.

Operations $M$ and $V$ occur at left edges on the lowest level of, respectively, inside the tree.
They pose a particular problem for two reasons. First PCRs cannot be directly written to, and 
a natural approach to reset them via \texttt{TPM\_PCR\_Reset} 
as a first step in $M$ or $V$ is problematic, since only PCRs above 16 
of a standard TPM can be reset, and only from the correct locality.
Thus it is necessary that enough PCRs are resettable and that they respond to the
locality in which the tree-formation software is executed as a trusted code.
Secondly, even after reset, the only operation which can modify a PCR, \texttt{TPM\_Extend},
does not directly copy a value into the register but truly executes~(\ref{eq:extend}) with
the existing value of the reset PCR, which is 160bit binary $0x00$ and the input value,
which yields a result different from the input value.
One option, which avoids exposing new commands directly writing to, or shifting values between PCRs,
would be to augment PCRs with a \texttt{reset} flag which indicates that they are in 
a pristine state after reset. Then, \texttt{TPM\_Extend} can be modified such that it directly writes 
into the PCR when this flag is \texttt{true}, and then sets it to \texttt{false}.
 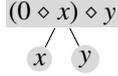
\begin{figure}[t]
 \centering
 \begin{tikzpicture}
 \node [style={rectangle}] {$(0\ex x)\ex y$}
 child {node {$x$}}
 child {node {$y$}};
 \end{tikzpicture}
 \caption{Hash configuration at one tree node depending on children values.}\label{fig:triangle}
 \end{figure}

Realising that $M$ and $V$ consistently occur at left edges of a tree, and only
if the right sibling is empty (\nil), and then deterministically
produce an outcome depending only on the two siblings involved, a third option would be to
deviate slightly from the definition of a Merkle hash tree. The correct configuration
of values in every elementary triangle in the SML tree would then be as shown in Figure~\ref{fig:triangle}.
That is $V$ or $M$ is modelled by \texttt{TPM\_PCR\_Reset} followed by \texttt{TPM\_Extend}
to obtain $0\ex x=H(0\|x)$ in the first step. The right sibling is then normally extended in
that register and the result written to the SML. See \ref{sec:useful-convention} for a
consistent treatment of \nil node values in intermediate stages and finalisation of a tree.

\subsection{\texttt{TPM\_Tree\_Extend}}\label{sec:texttttpm_tr}
The split TPM/software implementation of tree formation compromises on the security level of the
resulting root verification data register values.
It is preferable that tree-formed verification data is produced by a TPM-internal implementation 
of the proposed algorithms. For this, a TPM modification can work as follows.
The modified TPM exposes a new command \texttt{TPM\_Tree\_Extend} with the same input parameters
 as the usual \texttt{TPM\_Extend} command. 
 The TPM maintains flags for PCRs signifying which of them are currently designated tree roots, which are
 occupied and locked, and which are usable as intermediate $V$s by the algorithm. Furthermore,
 the TPM maintains the additional data mentioned in Section~\ref{sec:compl-tree-form}. In the simplest case,
 internal logic prevents concurrent use of more than one PCR for tree formation.
 While \texttt{TPM\_Extend} outputs only the update of the target PCR value, \texttt{TPM\_Tree\_Extend}
 returns a variable number $1,\ldots, d$ of updated verification register data values 
 in sequence such that they produce the natural order
 described in Section~\ref{sec:tree-form-proc}. 
 These return values are the output of the SML write operations of algorithms~\ref{alg:tree-form} 
 and~\ref{alg:tree-clean}.
 When $d$ values are returned, the receiver
 knows that this tree is exhausted and the corresponding root $V$ locked.
 Another option not considered here is for \texttt{TPM\_Tree\_Extend} to return all intermediate $V$s on
 each call.
\subsection{Experimental Performance}
We have
implemented the tree-formation procedure by introducing the new command  \texttt{TPM\_Tree\_Extend}
descibed in Section~\ref{sec:texttttpm_tr} in the TPM software 
emulation environment ethemba~\cite{ethemba} containing the TPM emulator~\cite{Strasser2004}.
For simplicity, the number of available PCRs in the emulator was increased and some (24) of them were 
exclusively assigned for use with the \texttt{TPM\_Tree\_Extend}
command.
Modifications to the software environment of the TPM emulator were also necessary.
In particular a new object \texttt{TcTpmTreeDigest} was defined in the TSS implementation
jTSS~\cite{jTSS} to receive the multiple return values from the tree formation algorithm.
The TPM-emulator-internal implementation
allows for an estimation of the relative size of the required code of 
the tree-formation algorithms. The  \texttt{TPM\_Tree\_Extend} command adds 271 logical (excluding comment 
and blank) lines of code (LoC) to the 21,756 logical LoC of the TPM emulator, i.e., about 1.3\%. 

To get an impression of the tree-formation algorithm's performance, 
random, 160-Bit values were generated and directly fed into \texttt{TPM\_Tree\_Extend} inside the jTSS. 
These random values were thus treated as (artificially generated) test-case  measurement values
and execution times were measured using the \texttt{SystemNanoTime} function. 
Samples were taken over $2^{15}$, $2^{17}$, and $2^{19}$ input measurement values, and over a number of 
individual runs, with each run starting from a freshly booted emulation system residing on a hardware platform with a 2,8GHz processor. The number of runs was chosen such that the total number of execution time measurements is $100\cdot 2^{19}$.
The average time, over all samples, of \texttt{TPM\_Tree\_Extend}
was $\approx218\mu$s, while that of \texttt{TPM\_Extend} was $\approx125\mu$s.
 \begin{figure}[t]
\centerline{\resizebox{0.5\textwidth}{!}{\includegraphics{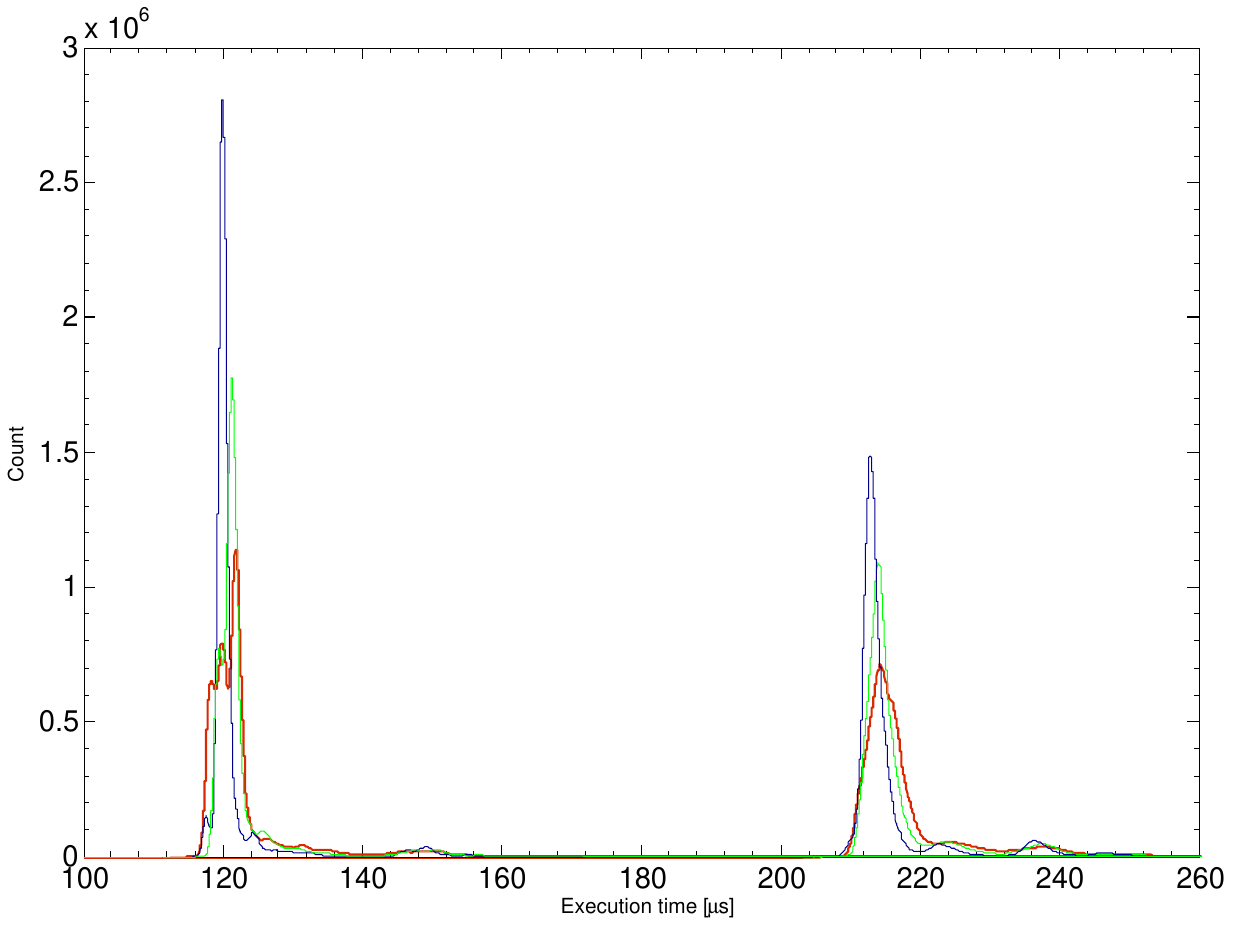}}} 
 \caption{Histogram counts of execution times of \texttt{TPM\_Extend} and \texttt{TPM\_Tree\_Extend} for $2^{15}$ (red) $2^{17}$ (green), and $2^{19}$ (blue)  input measurement values.}
 \label{fig:times}
 \end{figure}

The distribution of the execution times for  \texttt{TPM\_Extend} (left peaks) and 
 \texttt{TPM\_Tree\_Extend} (right peaks) is shown in Figure~\ref{fig:times}.
It is visible that, with the soft TPM emulation, the calculation of hash values is dominated
by data handling in the emulator and the jTSS environment. Essentially, \texttt{TPM\_Tree\_Extend}
has to handle, in particular during output to the SML, about double the data than \texttt{TPM\_Extend}. 
Some initial effects, probably due to
memory management, tend to wash out as the sample length increases.
\section{Related Work}\label{sec:related-work}
Verification of programs before loading and while booting was first mentioned 
in~\cite[Sections 6.2 and 6.3]{Lampson1992}, where a formalisation of the process 
is given and the concept of attestation appears.
Code authentication is among the primary goals of 
Trusted Computing~\cite{10.1109/MC.2003.1212691,10.1109/SP.1981.10016,10.1109/SP.1981.10020}.
Early work on protecting executed code by securing start up of a platform, such as 
Dyad~\cite{Tygar91dyad:a}, proposes hardware mechanisms to bootstrap trust in the 
host with secure coprocessors on standard PC hardware, and shows the first important 
applications of trusted platforms. 
Secure hardware must be involved in the secure bootstrap process. 
For instance, a secure coprocessor may halt the boot process if it detects an anomaly.
This assumes that the bootstrap ROM is secure. To ensure this, the system's address 
space could be configured such that the boot vector and the boot code are provided 
by a secure coprocessor directly or the boot ROM itself could be a piece of secure hardware. 
Regardless, a secure coprocessor verifies the system software (OS kernel, 
system related user-level software) by checking the software's signature against known values~\cite{Tygar91dyad:a}.

Tamper resistance of code has been considered by many researchers. A prototypical 
approach to the problem is rooting trust for program execution in hardware, such as the 
XOM (eXecute Only Memory~\cite{357005})  processor architecture, and the XOM Operating 
System~\cite{945463} building on it. This does not solve the problems of secure loading 
a program, and attesting to external entities.
AEGIS \cite{10.1109/SECPRI.1997.601317} shows 
secure boot on a PC. AEGIS uses a signed hash to identify each layer in the boot 
process, as does Terra~\cite{945464}, which can attest loaded components with a 
complete chain of certificates ending in attestation of virtual machines.

Existing TCG specifications define a bi-lateral remote attestation 
to verify the integrity of a platform remotely, by verifying the binary executables.
All executed code is measured when it gets loaded. The measurements are stored in PCRs as verification data, 
and the TPM attests to these data by signing them with a TPM protected key.
The verifier can, upon receipt of these metrics, decide if the platform can be considered trustworthy.
Since the whole configuration is transmitted and verified, the verifier needs to know all configurations of all machines.
Furthermore, binary attestation discloses the complete configuration and thus poses a privacy risk.
In~\cite{Property-Attestation_IBM_2004} and~\cite{Sadeghi2004, Chen2006}  ``property,'' respectively, 
``property-based attestation'' (PBA) are proposed.
PBA allows to assure the verifier of security properties of the verified 
platform without revealing detailed configuration data.
A trusted third party (TTP) is used to issue a certificate which maps the platform's configuration 
to the properties (in particular desired/undesired functionality) which can be fulfilled in this configuration.
The TPM can then, using a zero-knowledge proof, attest these properties to the verifier without disclosing 
the complete configuration. Essentially, PBA moves the infrastructural problem
of platform validation to a TTP, similarly to, but extending the role of, the TCG's privacy CA.

Another alternative is presented by the Nexus OS~\cite{1118613}
which builds on a minimal Trusted Computing Base (TCB) to establish strong 
isolation between user space and privileged programs.
Nexus has secure memory regions and monitoring and enforcement machines to
protect them. One application is to move device drivers into user space~\cite{Williams2008}.
Attestation by Nexus attaches descriptive labels to monitored programs and
thus allows for expressiveness similar to PBA, but intrinsic to the system.
Both the PBA concept, as well as the Nexus approach do not have means to validate a complex 
system comprised of a multitude of components, which furthermore shall be dynamically managed.
Both approaches are orthogonal to the present one, and could be combined
with it.

Hierarchical Integrity Management (HIM), see \cite{Cabuk2008a}, presents a dynamical framework for 
component-wise integrity measurement and policy-enabled management of platform components. 
Components and sub-components are related in HIM via dependency graphs, the most general 
structure that is useful for this purpose~\cite{Kuck1981,Ferrante1987}. But HIM is not aimed at (remote) platform 
validation and does not protect structured platform verification data in a PCR. Rather, 
it holds measurements in a global Component Configuration Register table of soft registers.

The main intended application of the hash trees introduced by Merkle for integrity 
protection of large datasets is in certificate management in a PKI. This yields long-term 
accountability of CAs, using Merkle trees~\cite{352604}, or authenticated search 
trees~\cite{603408}. Various groups have extended the use of hash trees to general 
long-term secure archiving for digital data~\cite{1083327,Kunz2007}. Corresponding 
data structures have been standardised in the so-called Evidence Record Syntax, 
by the IETF~\cite{rfc4998}.

A lot of research work has gone into the usage of hash trees for 
run-time memory protection. See Elbaz \textit{et al.}~\cite{Elbaz20091} and Hu \textit{et al.}~\cite{Hu2008} for a 
recent topical overviews over the state-of-the-art.
Typical systems employing hash trees for storage and memory 
protection~\cite{citeulike:5104772,Gassend03cachesand,10.1109/NCA.2004.1347805} 
separate a system into untrusted storage and a TCB.  
A program running on the TCB uses hash trees to maintain the integrity of data stored in untrusted storage, 
which can be, e.g., some easily accessible, bulk store in which the program regularly 
stores and loads data which does not fit into the TCB.
Gassend, \textit{et al.}~\cite{Gassend03cachesand} also propose to store the root of the
entire tree in an on-chip trusted register
of constant size, but keep all other nodes are in main memory or cache.

Regarding structuring of validation data beyond the simple linear verification data chain,
some attempts have been made.
The specifications of the TCG Infrastructure working group contain an 
approach to the problem of remotely validating a TCG Trusted Platform,
creating a hierarchy of components and sub-components~\cite{TCGIWGP1RA},
in a nested XML validation data structure. Although this data structure may
contain PCR values in the substructures, no binding by cryptographic means 
between those substructures, or substructures and parent structures, is foreseen.

On the academic side, Lo Presti~\cite{presti2007tree} proposed a Tree of Trust (ToT) concept and notation to represent 
a platform's structure. A ToT's nodes represent platform components, from TPM up to applications, 
annotated with trust and security statements. It can be used to assess the trust that should be put into the platform, 
or even to reorganise the platform according to certain constraints.
Another technical domain where the shortcomings of a merely
linear chain of trust becomes an urgent issue is virtualisation. Virtual machines are created and destroyed
dynamically on potentially many layers, resulting in ``a tree-like, dynamic structure of trust 
dependencies''~\cite[p.~6]{OpenTC_VTPM_Arch09}.
While the community has acknowledged that structured validation data is required to truly assess 
platforms' trustworthiness, a granular association of such tree-formed data hierarchies
to verification data (PCR values) is lacking.

The work most closely related to the present one is constituted by the proposal of Sarmenta, 
van Dijk, \textit{et al.}~\cite{sarmenta2006virtual},
to protect arbitrary memory objects via hash trees which in turn are protected by a root in TPM
non volatile memory. In~\cite{sarmenta2006virtual} a new TPM command \texttt{TPM\_ExecuteHashTree}
is introduced which allows to add, delete, and update so called \texttt{TPM\_COUNTER\_BLOB} objects,
and which issues a certificate, signed by an AIK, that attests to the successful
verification of that object's data with respect to the hash tree's root.
While this is a fully general method for handling arbitrary data sets in a TPM-protected hash tree,
it does not address the special problem of building the tree from sequentially arriving measurement
values maintaining the same security properties as the normal \texttt{TPM\_Extend} command.

A different usage of hash trees is proposed in~\cite{deng2006secure}, where it is shown how
they can support authentication of distributed code in Wireless Sensor Networks (WSN). Also
in WSN, data aggregation involving multiple nodes may be integrity protected using hash trees~\cite{1370623}.
Different from hash trees, another potential approach to make verification data searchable 
are Authenticated Append-only Skip Lists~\cite{Maniatis-thesis}, which
are sorted linked lists designed to allow fast 
lookup of the stored data elements by taking "shortcuts." However, trees are better 
suited for validation of a platform's state, in particular to efficiently determine 
the subset of components at the leaves failing validation.

Relative to the cited state-of-the-art, the contributions of the present paper are twofold.
First, we introduce a new method to generate a binary Merkle tree from component
measurement values using only a limited set of tamper-resistant verification data
registers, and existing capabilities of the TPM, i.e., the standard extend operation.
The algorithm is small enough to be executed within a TCB, in particular on-chip.
This part of our proposed method increases security of the generation of the root
of a hash tree, which in turn provides more security to the tree nodes. This
problem is, to the best of our knowledge, not considered in the literature. 
Second, we show how to exploit the tree structure for efficient validation with enhanced
diagnostic capabilities over common PCR values and SMLs, to increase security features 
of remote platform validation, and concurrently benefiting from the efficiency of tree-like
structures in the search for failure points. This use of tree-structured data for secure diagnostics, validation,
or attestation (all fields to which the proposed concepts apply), has also not been considered elsewhere, to 
the best of our knowledge.
\section{Discussion, Outlook, and Conclusion}\label{sec:disc-outl-concl}
Though hash trees are widely used, ours is the first proposal,
to the best of our knowledge, to use Merkle hash
trees to protect the integrity of the secure start up process of a trusted
platform in the same way as is traditionally done with PCRs.
We have demonstrated the efficiency and flexibility gains resulting
from using tree-formed verification data in platform validation.
This may be effective in particular in the remote validation and 
management of platforms via a network.
Given the small size and complexity of the tree-formation algorithm,
it seems possible to implement all these operations directly inside the TPM, if specifications 
are amended accordingly. This may or may not be a feasible approach
for future TPM generations,depending also on industry interests.
It could be that the specifications of the Mobile Trusted Module (MTM,~\cite{MTMREFARC})
have more flexibility toward incorporation of such advanced functionality.

Although the discussion framework for tree-formed verification data is that of Trusted Computing,
the main motivation for this work comes from a different angle.
Future, large scale deployments of networked devices, such as required in machine-to-machine communication
scenarios, require a solid device- and network-side, balanced and efficient trust 
infrastructure~\cite{AUS09B,CSSLM-VTM09}.
Security requirements are particularly high for devices loosely connected to networks and operating
semi-autonomously. Scenarios considered by the industry~\cite{TCM2M}, in particular regarding machine-to-machine (M2M)
communication of rather autonomously operating devices, always lead to the
requirement for remote integrity check, or validation, of a connecting device. To make validation
expressive, efficient, and secure, is a primary necessity. 
In fact, a variety of cryptographic checksum 
functions is available for checking code and data , and all variants obviously require that the integrity of the 
checksums for the "correct" data be maintained. The requirement for a centralised 
database of all software in all valid versions on the various machines is a significant 
management problem, in need of an efficient solution. A ``work-around'' which, however, incurs a loss of granularity in
platform validation, is obviously to check only system parts considered security-sensitive, such as a secure OS loader or
a hypervisor.
On the other hand, the proposed approach to validation using tree-formed data promises increased scalability for such use cases.
The natural hierarchy of trees suggests natural combinations with analogously hierarchical network architectures, e.g.,
where large numbers of (rather uniform or representing a limited number of device classes) M2M devices are connected 
via gateways to networks. According validation architectures and procedures are described in~\cite{IJDTIS-TPVM10}.

A further variant of validation with tree-formed data can save communication overhead by straightforwardly turning
the tree-traversal of diagnostic validation into an \textit{interactive} procedure between platform and validator.
This allows to send only the data actually required for traversal, level by level, to the validator, and thus
restrict communication to a minimum. This is described in~\cite{InteractivePVM2012}.
A demonstration implementation in which an emulated Android platform is the validation target,
shows the feasibility of the concept of interactive remote validation with tree-formed verification data
This demonstration can be viewed at~\cite{ethemba}.

With regard to generalisations, trees are certainly not the most general structures 
for which integrity protection using cryptographic digests can be applied. 
For instance, some researchers have extended hashes to provide identification of
directed graphs~\cite{Portegys2007}. Others have applied variations of one-way functions, e.g., multi-set
hashes~\cite{Clarke03incrementalmultiset} to uniquely identify complex data structures such as RDF graphs~\cite{Sayers2004}.
Along these lines, generalisation of tree-formed verification data to, for instance,
directed acyclic graphs, and dependence graphs~\cite{Kuck1981,Ferrante1987} can be conceived.
While potentially interesting for complex platform management and protection tasks,
every such generalisation would incur increased complexity and lose the efficiency of
binary trees for validation.

Also on the device side, the hierarchical structure of trees suggests improvements for validation.
A natural one is to define various assurance levels for verification data beyond the protected/unprotected
dichotomy. For instance, higher levels of verification data trees could be kept in software (e.g.\ 
using virtualisation) or hardware protected regions of memory. Such architectural extensions may also
allow to use the tree-formed approach for different kinds of start up procedures than ``cold boot'' only.

The single command extension of the TPM integrity measurement functionality,
$\texttt{TPM\_Tree\_Extend}$ proposed above is, however, only the starting point
of a flexible, TPM-based tree verification data management architecture.
In particular it would be desirable to enable secure updates of subtree roots,
for instance for dynamic platform management,
and ultimately to quote an inner node of a tree-formed SML with the same security
assertions as $\texttt{TPM\_Quote}$ provides to a remote validator for a PCR value.
Finally, a desired feature is the ability to certify subtrees via a TTP, 
as discussed in Section~\ref{sec:extens-effic-augm}. Such extensions are described in~\cite{DBLP:TFV2}.
\section*{Acknowledgements}
This work was funded by InterDigital, Inc. 
Special thanks go to
Lawrence Case,
Bob DiFazio 
David Greiner,
Louis Guccione, 
Dolores Howry,
Michael V. Meyerstein,
and
Sudhir Pattar, 
for many useful discussions and comments.
The authors are grateful to the anonymous referees, 
whose comments and suggestions  helped to improve the
article's structure and contents. 
\appendix
\section{A Useful Convention}\label{sec:useful-convention}
In many cases, the hash tree stored in the SML will be incomplete, i.e., contain
empty leaves and inner nodes. In the continuous measurement process, such nodes, with
value denoted \nil, are treated procedurally by the operations $M$ and $V$ 
(see Section~\ref{sec:compl-tree-form}) which means that right \nil siblings
are ignored. This happens in lines~\ref{ln:M} and~\ref{ln:shifta} of Algorithm~\ref{alg:tree-form}
for intermediate stages of tree formation, and in line~\ref{ln:shiftb}
of Algorithm~\ref{alg:tree-clean} at completion of the tree after the last measurement.

Generally, i.e., transgressing the restrictions of a standard TPM, 
it may be useful to assume that \nil is a two-sided unit for the operation \ex, i.e.,
\begin{equation*}
  \label{eq:exunit}
  x\ex\nil = \nil\ex x = x, \text{and } \nil\ex\nil=\nil.
\end{equation*}
This convention is in accordance with the rule for validation of node configuration
Fig.\ref{fig:nodeclasstamper}~(d). It is a 
re-interpretation of the usual extend operation and can also be used to eliminate
the operations $M$ and $V$ in the algorithms' formulations. 
Namely, $M$ and $V$ can be replaced by a reset of a register $V$ to \nil followed
by the operation $V\gets V\ex m$, respectively $V\gets V\ex V'$.

For the implementation of this convention, 
we may assume that \nil is to be represented
as an additional flag of PCR registers,  and the inputs and output of \ex.
For a PCR, the \nil flag is set by a particular reset command.
When \nil is encountered as the input
of an extend operation to a PCR, then logic of the TSS, or a TPM modification, may
prevent execution of the hash operation (\ref{eq:extend}) and write to the PCR directly. 
\section{The Expected Number of Bad Inner Nodes}\label{sec:calc-expect-numb}
The problem under consideration is that of bi-colouring (bad vs.\ good inner nodes) 
of a binary tree generated
by a random, i.i.d.\ choice of leaves and colouring of the path connecting it to the root.
Random choices of such leaves and paths is equivalent to random choices of i.i.d.\ bit
strings of length $d$. We first calculate the expected number $E_k^N$ of coloured leaves 
after $k$ choices from the set of $N=2^d$ leaves. Recursively, $E_0^N=0$, and 
\[
E_{k+1}^N=E_k^N\frac{E_k^N}{N}+(E_k^N+1)\frac{1-E_k^N}{N}=
1+E_k^N-\frac{E_k^N}{N}.
\]
Solving this obtains
\[
E_k^N=N\left(1-\left(1-N^{-1}\right)^k\right).
\]
Since all substrings of the chosen bit-strings are statistically independent the
same argument applies to inner nodes at all levels $d-1,\ldots,0$. Thus, the expected number
of coloured inner nodes is obtained by summation
\[
E_k^\text{inner}=\sum_{\ell=0}^{d-1} E_k^{2^\ell}.
\]
It remains to find the expected number of choices $k$ which corresponds to a certain 
expected number $E_k^N=fN$ of coloured leaves, where $0\leq f\leq1$ is a target fraction of leaves.
Solving this equation for $k$ yields
\[
k=\frac{\ln(1-f)}{\ln(1-2^{-d})},
\]
where $N=2^d$ was inserted. From this, the expected number of bad inner nodes in dependency
of $f$, $E^{\text{inner}}(f)$, can be calculated.


\bibliographystyle{elsarticle-num}
\bibliography{CoSe_TFV1}


\end{document}